\documentclass[multphys,vecphys]{svmult}

\usepackage{makeidx}         
\usepackage{graphicx}        
\usepackage{multicol}        
\usepackage[bottom]{footmisc}

\usepackage{floatflt}

\newcommand{\calastrom}{\mbox{Str\"omgren~}}
\newcommand{\calaomc}{\mbox{$\omega$ Cen~}} 

\makeindex             

\begin{document}

\title*{Relative and absolute calibration for multi-band data collected
with the 2.2m ESO/MPI and 1.54m Danish Telescopes}
\titlerunning{Calibration of multi-band data collected with ESO Telescopes} 

\author{A. Calamida\inst{1}\and C. E. Corsi\inst{1}\and G. Bono\inst{1}, P. B. Stetson\inst{2},
L. M. Freyhammer\inst{3}\and R. Buonanno\inst{4}}
\authorrunning{A. Calamida}

\institute{INAF-Osservatorio Astronomico di Roma, Via Frascati 33, 00040,
Monte Porzio Catone, Italy,
\texttt{calamida@mporzio.astro.it}
\and DAO, HIA-NRC, 5071 W. Saanich Road, Victoria, BC V9E~2E7, Canada
\and Univ. of Central Lancashire, Preston PR1 2HE, UK 
\and Univ. di Roma Tor Vergata, Via della Ricerca Scientifica 1, 00133 Rome, Italy}
%
%
\maketitle

\begin{abstract}
We present the strategies adopted in the relative and absolute calibration of two
different data sets: $U,B,V,I$-band images collected with the Wide Field Imager (WFI)
mosaic camera mounted on the 2.2m ESO/MPI Telescope and $u,v,b,y$ \calastrom images
collected with the 1.54m Danish Telescope (ESO, La Silla). In the case of the WFI camera
we adopted two methods for the calibration, one for images collected before 2002, with the 
ESO filters $U/38_{ESO841}$ and $B/99_{ESO842}$, and a different one for data
secured after 2002, with the filters $U/50_{ESO877}$ and $B/123_{ESO878}$. 
The positional and color effects turned out to be stronger for images collected with
the old filters. The eight WFI chips of these images were corrected one by one, while in 
the case of images secured with the new filters, we corrected the entire mosaic in a
single step. In the case of the Danish data set, we compared point-spread 
function (PSF) and aperture photometry for each frame, finding a trend in both the X and Y directions
of the chip. The corrections resulted in a set of first and second order polynomials
to be applied to the instrumental magnitudes of each individual frame as a function
of the star position.
\end{abstract}

\section{WFI data set}
\subsection{Observations and data reduction}
Data have been retrieved from the ESO archive and include 8 $U$, 39 $B$, 51 $V$,
and 26 $I$ images of \calaomc. Data include both shallow and relatively deep images,
with exposures times ranging from 1 to 300s for the $B, V, I$ bands, and from 300 to
2400s for the $U$ band, and were collected in several observing runs ranging from 1999 to 2003.
During this period two filters were changed: data secured before 2002 were collected with the
filters $U/38_{ESO841}$ and $B/99_{ESO842}$, while later ones with the filters
$U/50_{ESO877}$ and $B/123_{ESO878}$. These data were obtained in good seeing conditions, 
and indeed the mean seeing ranges from ~0.6" for the $I$ band to ~1.1" for the
$U$ band.
We accurately selected the best PSF stars uniformly distributed across each chip. 
A moffat analytical function linearly variable on the chip was assumed for the PSF.
The data were reduced with DAOPHOT {\small IV}/ALLFRAME (Stetson 1994). 
\subsection{Relative and absolute calibration}
We are interested in providing accurate relative calibration of individual chips of
the WFI mosaic camera because the occurrence of positional effects might cause a spurious
color broadening of key evolutionary features such as the Red-Giant Branch (RGB) and the 
Main-Sequence Turn-Off. Moreover, we need to provide an accurate absolute
calibration of our data in order to compare theory with observations. The plausibility of 
this comparison relies on the accuracy of the absolute zero-point of individual bands. 
This is a fundamental requirement for the distance modulus, and in turn for the absolute ages.
Recent findings (Corsi et al. 2003; Koch et al. 2004) based on photometric data collected 
with the WFI indicate that the eight CCD chips might be affected by positional effects involving
zero-point errors of the order of several hundredths of magnitude. This subtle effect could
be due to scattered light. It seems that telescopes equipped with mosaic
cameras and focal reducers may present this problem (Manfroid et al. 2001, Manfroid \& Selman 2001).
\begin{floatingfigure}[l]{0.45\textwidth}
\centering
\includegraphics[height=5.cm,width=5.5cm]{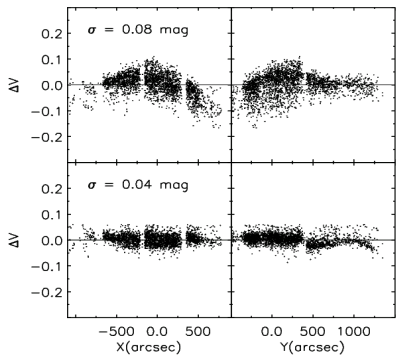}
\caption{Residuals, $\Delta mag = mag_S - mag_{PSF}$, in the $V$ band, plotted versus
the X, Y position on the chip, before (top) and after (bottom) the positional
corrections were applied, for an image collected in 2002.}
\end{floatingfigure}

In order to correct the positional effects of the WFI mosaic camera and to perform an accurate
absolute calibration of \calaomc data set, we followed these steps:\\
$\bullet$ obtain a set of local standard stars for \calaomc;\\
$\bullet$ identify all the trends of PSF magnitudes compared to standard
magnitudes versus position on the frame and correct them;\\
$\bullet$ estimate the calibration curves.\\
We thus made use of a set of new multi-band ($U,B,V,I$) local standard stars for \calaomc
 (Stetson et al. 2007). This star list has been selected in photometric accuracy ($\sigma \leq$ 0.03 mag) 
and in 'separation index' ($sep \geq$ 2.5). We ended up with a catalog of
$\sim 3\times 10^4$ local standard stars. The sky area covered by these 
stars is $\sim$ 37'$\times$40', and includes a substantial fraction of our WFI data. 
We thus applied two methods: in the case of images collected before 2002 ($U/38_{ESO841}$ and
$B/99_{ESO842}$ filters), the eight chips were corrected one by one, while for images secured starting from 2002
($U/50_{ESO877}$ and $B/123_{ESO878}$ filters) we corrected the entire mosaic in a single step.
The color terms for these filters have a very steep slope, therefore, we decided
to estimate a first color curve and apply it. We then studied the residuals,
$\Delta mag = mag_S - mag_{PSF}$, where $S$ stands for Standard, as a function 
of the X and Y position on the chip (or mosaic, see Fig. 1). Once corrected for the positional effects,
we estimated the color term once again and applied it for the absolute 
zero-point calibration. This strategy relies on the assumption that the two corrections are independent, and there 
is no reason why this should not be the case. 
Fig. 1 shows the magnitude residuals in the $V$ band plotted
versus the X and Y positions on the frame, before and after the corrections were applied to
an image collected in 2002. 
The trend with the position is clear, non-linear, and stronger in the four
outermost chips. A small residual trend on the position is still present in the
external regions, where the lack of standard stars makes it difficult to
estimate the corrections. 

\begin{floatingfigure}[l]{0.45\textwidth}
\centering
\includegraphics[height=5cm,width=5cm]{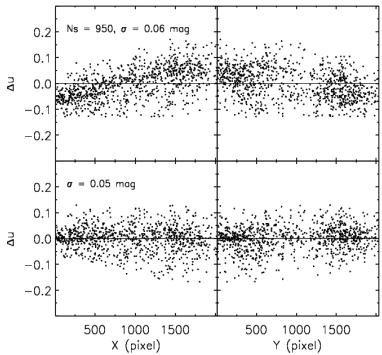}
\caption{Residuals, $\Delta mag = mag_{AP} - mag_{PSF}$, in the $u$ band, plotted versus
the X, Y position on the chip, before (top) and after (bottom) the
positional corrections were applied.}
\end{floatingfigure}

We thus corrected the positional effects by fitting the $\Delta mag$ vs $X/Y$ with second, 
third or fourth order polynomials, and we then estimated the calibration curves for each set.
We adopted a first order polynomial to calibrate the $V$ and the $I$ bands as a function of
the instrumental $V-I$ color. In the case of the $B$ band, we used a first order polynomial to
calibrate the new filter, $B/123_{ESO878}$, and a third order one for the old filter,
$B/99_{ESO842}$, as a function of the instrumental $B-V$ color.
Particular attention has been paid to the $U$ band calibration, either in the case of the old
filter, $U/38_{ESO841}$, as well in the case of the new one, $U/50_{ESO877}$. Having applied
the positional corrections to each set, we estimated a first color curve, a fourth order
polynomial for $U$ vs $U-I$. After applying this calibration curve, the $U$ 
magnitudes still showed a residual trend as a function of the $U$ magnitude and
the $U-I$ color. We thus corrected these trends with first and sixth order 
polynomials, respectively. The reason of these very complicated trends with colors is probably due to the shape of the old $B$ and $U$ filters,
which is quite different from the shape of the standard Johnson filters.

\section{Danish data set}
A set of 110 $u,v,b,y$ \calastrom images centered on \calaomc were collected 
in 1999, with the 2048$\times$2048 pixel Ford-Loral CCD mounted on the
Danish Telescope. Data have been reduced with DAOPHOT {\small IV}/ALLFRAME, using an IDL
procedure to accurately select the best PSF stars across each frame. We found that the $b$ and
the $y$ frames present faint spurious reflected images of bright stars. This effect causes
a systematic decrease in the flux measured by the PSF of bright stars and a systematic increase
in the flux of the faint neighbor stars.
The effect strongly affects the photometry in crowded fields, resulting in a
larger dispersion on the color-magnitude diagrams (CMDs), in particular in
$b$ versus $b-y$. 
Furthermore, we investigated the 
presence of magnitude trends with X, Y position on the chip, comparing PSF and
aperture photometry for each frame. The resulting effect changes from frame to frame, depending
on the photometric band, and the magnitude difference from one edge to the other of the image can be as high as
0.15 mag. The result is a set of first and second order polynomials that correct the
magnitudes of each individual frame as a function of the X and Y coordinates. 
Fig. 2 shows the magnitude residuals, $\Delta mag = mag_{AP} - mag_{PSF}$, for
the $u$ band,
before and after applying the corrections. 
\section{Conclusions}
\begin{floatingfigure}[r]{0.5\textwidth}
\centering
\includegraphics[height=6.cm,width=5.8cm]{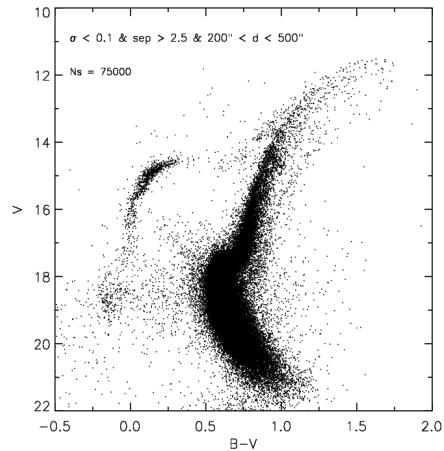}
\caption{$V, B-V$ CMD for the entire WFI corrected catalog of \calaomc.}
\end{floatingfigure}
We have presented the relative and absolute calibration strategies applied to extended photometric
catalogues obtained with the 2.2m MPI and the 1.54m Danish ESO Telescopes. 
The photometry was performed with DAOPHOT {\small IV}/\\ALLFRAME in both cases, using a linearly variable
PSF across the images. In spite of the attention paid in selecting PSF stars uniformly
distributed in the frame, we found a systematic error depending on the position of the stars
in the image, when comparing instrumental to standard magnitudes.
This effect might be due to
problems in adapting the analytical PSF to variations across the image or to intrinsic problems
of the optical system. Therefore, in order to perform an accurate relative and absolute
calibration it is useful to check for the presence of a positional dependence of photometric
errors. This is particularly recommended when dealing with telescopes equipped with focal
reducers and mosaic cameras.

%
%

%
%



\printindex
\end{document}